\documentclass[twocolumn,aps,prd]{revtex4-2}
\usepackage{graphicx,amssymb}
\usepackage[utf8]{inputenc} 
\PassOptionsToPackage{obeyspaces,spaces,hyphens}{url}\usepackage[urlcolor=blue,colorlinks,breaklinks=true]{hyperref}
\usepackage{amsmath}
\usepackage{float}
\usepackage{bm}
\DeclareUnicodeCharacter{2212}{-}
\begin{document}

\title{Distance-duality in theories with a nonminimal coupling to gravity}

\author{R.P.L. Azevedo}
\email[Electronic address: ]{rplazevedo@astro.up.pt}
\affiliation{Departamento de F\'{\i}sica e Astronomia, Faculdade de Ci\^encias, Universidade do Porto, Rua do Campo Alegre 687, PT4169-007 Porto, Portugal}
\affiliation{Instituto de Astrof\'{\i}sica e Ci\^encias do Espa{\c c}o, Universidade do Porto, CAUP, Rua das Estrelas, PT4150-762 Porto, Portugal}

\author{P.P. Avelino}
\email[Electronic address: ]{pedro.avelino@astro.up.pt}
\affiliation{Departamento de F\'{\i}sica e Astronomia, Faculdade de Ci\^encias, Universidade do Porto, Rua do Campo Alegre 687, PT4169-007 Porto, Portugal}
\affiliation{Instituto de Astrof\'{\i}sica e Ci\^encias do Espa{\c c}o, Universidade do Porto, CAUP, Rua das Estrelas, PT4150-762 Porto, Portugal}

\date{\today}
\begin{abstract}

We show that gravitational theories with a nonminimal coupling (NMC) to the matter fields lead to a violation of Etherington's distance-duality relation, which relates the luminosity and angular-diameter distances. We derive constraints on power law and exponential NMC models using existing measurements of type Ia supernovae and baryon acoustic oscillations throughout the redshift range $0<z<1.5$. These complement previous constraints derived from cosmic-microwave background radiation and big bang nucleosynthesis data.

\end{abstract}
\maketitle

\section{Introduction}
\label{sec:intr}

General relativity (GR) remains the most successful theory of gravitation, passing many experimental hurdles, but it nevertheless requires the addition of dark energy and dark matter to explain the accelerated expansion of the Universe and the nontrivial dynamics of cosmological structures such as galaxies and clusters thereof \cite{Bertone2005,Carroll2001}. However, the existence of these dark components is merely inferred, as direct observation has not yet been possible. Thus, alternatives have been proposed to explain the observed dynamics without the need for dark energy and/or dark matter. These include theories with more complex geometric terms such as $f(R)$ and $f(R, R_{\mu\nu}, R_{\mu\nu\alpha\beta})$, and theories featuring a nonminimal coupling (NMC) between geometry and matter, such as $f(R,\mathcal{L}_\text{m})$ theories \cite{Nojiri2004,Allemandi2005,Bertolami2007,Sotiriou2008,Harko2010,Harko2011} (see, however, \cite{Clowe2004,Markevitch2004}).

NMC theories of gravity feature significant changes not only to cosmological dynamics and thermodynamics, but also to energy-momentum conservation \cite{Aviles2011,Avelino2018, Azevedo2018a,Azevedo2019a,Avelino2020a,Azevedo2020}. These changes depend on the Lagrangian of the matter fields which appears explicitly in the equations of motion and, therefore, the use of the correct Lagrangian is crucial in order to be able to derive useful constraints on NMC gravity. In previous work $\mathcal{L}_\text{m}=-\rho$ or $\mathcal{L}_\text{m}=p$ have been suggested as the on-shell Lagrangian of a perfect fluid with proper energy density $\rho$ and pressure $p$ \cite{Nesseris2009,Bertolami2012,Azizi2014,Ribeiro2014,Silva2018}. However, the correct on-shell Lagrangian for a fluid composed of classical particles of fixed rest mass and structure has been shown to be given by the trace of the energy-momentum tensor of the fluid $\mathcal{L}_\text{m} = T$ \cite{Avelino2018a,Avelino2018}. This is expected to be an excellent approximation in the case of baryons, dark matter, and photons --- in the latter case, the zero rest mass limit should be considered. However, it does not apply to dark energy (or to any fluid with an equation of state parameter $w$ outside the interval $0 \leq w \leq 1/3$).

Etherington's relation, also known as the distance-duality relation (DDR), directly relates luminosity and angular-diameter distances in GR, where they differ only by a specific function of the redshift. It has recently come into focus given the possibility of performing more accurate tests of Etherington's relation  with new cosmological surveys of type Ia supernovae (SnIa) and baryon acoustic oscillations (BAO)\cite{Bassett2004,Ruan2018,Xu2020,Martinelli2020,Lin2021,Zhou2021}, as well as observations from Euclid \cite{Laureijs2011,Astier2014} and gravitational wave observatories \cite{Cai2017}. In this work we derive the impact of NMC theories on the DDR, and use the most recent data available to impose constraints on a broad class of NMC models.

Throughout this paper we use  units such that $c=1$, where $c$ is the value of the speed of light in vacuum. We adopt the metric signature $(-,+,+,+)$, and the Einstein summation convention shall be used as usual. We use Greek and Latin indices for four- and three-dimensional quantities, respectively.

\section{Nonminimally Coupled Gravity}
\label{sec:model}

The action
 	\begin{equation}
	\label{eq:action}
	S=\int \sqrt{-g} \left[\kappa f_1(R) + f_2(R)\mathcal{L}_\text{m}^\text{(off)}\right]\,,
\end{equation}
describes a wide class of $f(R)$-inspired NMC gravity models. Here, $\kappa=(16\pi G)^{-1}$, $G$ is Newton's gravitational constant, $g$ is the determinant of the metric $g_{\mu\nu}$, $\mathcal{L}_\text{m}^\text{(off)}$ is the off-shell Lagrangian of the matter fields, and $f_1(R)$ and $f_2(R)$ are generic functions of the Ricci scalar $R$. This class of models allows for a broad range of cosmological dynamics while showcasing the effects of the NMC on the dynamics of the matter fields, and is widely used in the existing literature. Notice that GR is recovered if $f_1(R)=R$ and $f_2(R)=1$. Extremizing the action with respect to the metric one obtains the equations of motion of the gravitational field
\begin{equation}
	\label{eq:eqmotion}
	F G_{\mu\nu}=\frac{1}{2} f_2 T_{\mu\nu}+\Delta_{\mu\nu}F+\frac{1}{2}\kappa f_1 g_{\mu\nu}-\frac{1}{2} RFg_{\mu\nu}\, ,
\end{equation}
where $G_{\mu\nu}=R_{\mu\nu}-\frac{1}{2} g_{\mu\nu} R $ is the Einstein tensor, $R_{\mu\nu}$ is the Ricci tensor, $\Delta_{\mu \nu} \equiv \nabla_\mu \nabla_\nu - g_{\mu \nu} \Box$, $\Box \equiv \nabla^\mu \nabla_\mu$,  
\begin{equation}
	\label{eq:F}
	F=\kappa f'_1(R)+f'_2(R)\mathcal{L}_\text{m}\,,
\end{equation}
a prime denotes a derivative with respect to the Ricci scalar, $\mathcal{L}_\text{m}$ is the on-shell Lagrangian of the matter fields, and the energy-momentum tensor has the usual form
\begin{equation}
	\label{eq:energymom}
	T_{\mu\nu}=-\frac{2}{\sqrt{-g}}\frac{\delta(\sqrt{-g}\mathcal{L}_\text{m}^\text{(off)})}{ \delta g^{\mu\nu}}\,.
\end{equation}

Taking the covariant derivative of Eq.~\eqref{eq:eqmotion} and using the Bianchi identities one obtains the following relation
\begin{equation}
	\label{eq:noncons}
	\nabla^\mu T_{\mu\nu}=\frac{f'_2}{ f_2}(g_{\mu\nu}\mathcal{L}_\text{m}-T_{\mu\nu})\nabla^\mu R\, ,
\end{equation}
which implies that the form of the on-shell matter Lagrangian directly affects not only energy-momentum conservation, but also particle motion \cite{Bertolami2008,Avelino2018, Avelino2020a}. 

The on-shell Lagrangian for a single-point particle with action
\begin{equation}
	\label{eq:act_part}
	S=-\int d\tau\,m \,,
\end{equation}
and energy-momentum tensor
\begin{equation}
	\label{eq:emt_part}
	T^{\mu\nu}=\frac{m}{\sqrt{-g}}\int d\tau \,u^\mu u^\nu \delta^4 (x^\sigma-\xi^\sigma(\tau)) \,, 
\end{equation}
is given by the trace of the energy-momentum tensor (see, for example, \cite{Avelino2020a,Ferreira2020a}) 
\begin{equation}
	\label{eq:part_lag}
	\mathcal{L}_\text{m}=T=-\frac{m}{\sqrt{-g}}\int d\tau \, \delta^4 (x^\sigma-\xi^\sigma(\tau)) \,,
\end{equation}
where $\delta^4 (x^\sigma-\xi^\sigma(\tau))$ denotes the four–dimensional Dirac delta function, $\xi^\sigma(\tau)$ represents the particle worldline, $\tau$ is the proper time, $u^\mu$ are the components of the particle four-velocity ($u^\mu u_\mu=-1$), and $m$ is the proper particle mass. An alternate derivation of this result, $\mathcal{L}_\text{m}=T$, for an ideal gas is given in \cite{Avelino2018a,Avelino2018} (see also \cite{Ferreira2020a} for a detailed discussion of the appropriateness of the use of different Lagrangians to describe various components of the cosmic energy budget). The covariant derivative of Eq. \eqref{eq:emt_part} may thus be written as
\begin{equation}
	\label{eq:cov_emt_part}
	\nabla_\nu T^{\mu\nu}=\frac{m}{\sqrt{-g}}\int d\tau\, (\nabla_\nu u^\mu) u^\nu \delta^4 (x^\sigma-\xi^\sigma(\tau)) \,, 
\end{equation}
and using Eqs. \eqref{eq:emt_part}, \eqref{eq:part_lag} and \eqref{eq:cov_emt_part} in Eq. \eqref{eq:noncons} one obtains the equation of motion of a point particle
\begin{equation}
	\label{eq:geodesic}
	\frac{du^{\mu} }{ ds}+\Gamma^\mu_{\alpha\beta} u^\alpha u^\beta = \mathfrak{a}^\mu \,,
\end{equation}
where $\mathfrak{f}^\mu = m\mathfrak{a}^\mu$ is a momentum-dependent four-force, given by
\begin{equation}
	\label{eq:extraforce}
	\mathfrak{f}^\mu=-m\frac{f'_2}{ f_2}h^{\mu\nu}\nabla_\nu R \, ,
\end{equation}
and $h^{\mu\nu}=g^{\mu\nu}+u^{\mu}u^{\nu}$ is the projection operator.

Here we shall consider a flat, homogeneous and isotropic Friedmann-Lemaître-Robertson-Walker (FLRW) universe with line element
\begin{equation}
	\label{eq:metric}
	ds^2=-dt^2+a^2(t)d\bm{r}\cdot d\bm{r}\,,
\end{equation} 
where $a(t)$ is the scale factor, $t$ is the cosmic time, and $\bm{r}$ are cartesian coordinates. Solving Eqs. \eqref{eq:geodesic} and \eqref{eq:extraforce}  in this metric one finds that the components of the three-force on particles $\bm{\mathfrak{f}}=d\bm{p}/dt$ are given by \cite{Avelino2020a}
\begin{equation}
	\label{eq:3force}
	\mathfrak{f}^i = -\frac{d\ln(af_2)}{dt}p^i \,,
\end{equation}
where $p^i$ are the components of the particle's three-momentum $\bm{p}$, which therefore evolve as
\begin{equation}
	p^i \propto (a f_2)^{-1}\,. \label{eq:momev}
\end{equation}
See \cite{Avelino2018} for an alternate derivation of this result in the case of solitons with fixed mass and structure. Assuming the zero-mass limit for photons, this implies that their frequency $\nu$, and their energy $E$, no longer evolve with $a^{-1}$ as in GR. The equation for the evolution of $E$ and $\nu$ in NMC gravity is instead given by
\begin{equation}
	\label{eq:photon_freq}
	E\propto\nu\propto \frac{1}{af_2} = \frac{1+z}{f_2} \, ,
\end{equation}
where $z$ is the redshift, and we have set $a=1$ at the present time. This alteration to the dynamics of photons, associated to the additional factor of $f_2^{-1}$, leads to stringent constraints on NMC theories from both cosmic microwave background (CMB) and big bang nucleosynthesis (BBN) data \cite{Avelino2018}.

We consider that the universe is filled by a collection of perfect fluids, with an energy-momentum tensor of the form
\begin{equation}\label{eq:pf_emt_nmc}
	T^{\mu\nu}=(\rho+p)U^\mu U^\nu + p g^{\mu\nu}\,,
\end{equation}
where $\rho$ and $p$ are the proper density and proper pressure of the fluid, respectively, and $U^\mu$ are the components of the four-velocity of a fluid element, satisfying $U_\mu U^\mu = -1$. Using Eqs. \eqref{eq:metric} and \eqref{eq:pf_emt_nmc} in the field equations \eqref{eq:eqmotion}, we obtain the modified Friedmann equation (MFE)
\begin{equation}\label{eq:fried-f1f2-1}
	H^2=\frac{1}{6F}\left[FR- \kappa f_1+f_2\rho-6H\dot{F}\right]\,,
\end{equation}
and the modified Raychaudhuri equation (MRE)
\begin{equation}\label{eq:ray-f1f2-1}
	2\dot{H}+3H^2=\frac{1}{2F}\left[FR-\kappa f_1-f_2 p-4H\dot{F}-2\ddot{F}\right] \,,
\end{equation}
where a dot represents a derivative with respect to the cosmic time, $H\equiv\dot{a}/a$ is the Hubble factor, and $F$ is defined by Eq. \eqref{eq:F} with  $\mathcal{L}_\text{m}=T=3p-\rho$.

\section{The distance-duality relation}

The luminosity distance $d_\text{L}$ of an astronomical object relates its absolute luminosity $L$, i.e. its radiated energy per unit time, and its energy flux at the detector $l$, so that they maintain the usual Euclidian relation
\begin{equation}
	\label{eq:apparent luminosity}
	l = \frac{L}{4\pi d_\text{L}^2}\,,
\end{equation}
or in terms of the luminosity distance
\begin{equation}
	\label{eq:lum_dist_1}
	d_\text{L} = \sqrt{\frac{L}{4\pi l}}\,.
\end{equation}
Over a small emission time $\Delta t_\text{em}$ the absolute luminosity can be written as
\begin{equation}
	\label{eq:abs_lum_1}
	L = \frac{N_{\gamma,\text{em}} E_\text{em}}{\Delta t_\text{em}}\,,
\end{equation}
where $N_{\gamma,\text{em}}$ is the number of emitted photons and $E_\text{em}$ is the average photon energy. An observer at a coordinate distance $r$ from the source will, however, observe an energy flux given by
\begin{equation}
	\label{eq:app_lum}
	l = \frac{N_{\gamma,\text{obs}}E_\text{obs}}{\Delta t_\text{obs} 4\pi r^2}
\end{equation}
where $N_{\gamma,\text{obs}}$ is the number of observed photons and $E_\text{obs}$ is their average energy.

Note that while the number of photons is conserved, $N_{\gamma,\text{obs}} = N_{\gamma,\text{em}}$, the time that it takes to receive the photons is increased by a factor of $1+z$, $t_\text{obs}= (1+z)t_\text{em}$, and as per Eq. \eqref{eq:photon_freq}, their energy is reduced as 
\begin{equation}
	\label{eq:energy_obs}
	E_\text{obs} = \frac{E_\text{em}}{1+z}\frac{f_2(z)}{f_2(0)} \,,
\end{equation}
where $f_2(z)=f_2[R(z)]$ and $f_2(0)=f_2[R(0)]$ are respectively the values of the function $f_2$ at emission and at the present time. The distance $r$ can be calculated by just integrating over a null geodesic, that is
\begin{align}
	\label{eq:null geodesic}
	ds^2 &= - dt^2 + a(t)^2dr^2 = 0\nonumber \\
	\Rightarrow dr &= -\frac{dt}{a(t)} \nonumber \\
	\Rightarrow  r &= \int_{t_\text{em}}^{t_\text{obs}} \frac{dt}{a(t)}= \int_0^z \frac{dz'}{H(z')} \,,
\end{align}
Using Eqs. \eqref{eq:abs_lum_1}, \eqref{eq:app_lum}, \eqref{eq:energy_obs} and \eqref{eq:null geodesic} in Eq. \eqref{eq:lum_dist_1}, we finally obtain
\begin{equation}
	\label{eq:lum_dist}
	d_\text{L} = (1+z)\sqrt{\frac{f_2(0)}{f_2(z)}}\int_0^z \frac{dz'}{H(z')} \, .
\end{equation}
In the GR limit $f_2=\text{const.}$, and we recover the standard result

The angular-diameter distance $d_\text{A}$, on the other hand, is defined so that the angular diameter $\theta$ of a source that extends over a proper distance $s$ perpendicularly to the line of sight is given by the usual Euclidean relation
\begin{equation}
	\label{eq:angle}
	\theta=\frac{s}{d_\text{A}}\,.
\end{equation}
In a FLRW universe, the proper distance $s$ corresponding to an angle $\theta$ is simply
\begin{equation}
	\label{eq:prop_dist}
	s= a(t)r\theta = \frac{r\theta}{1+z}\,,
\end{equation}
where the scale factor has been set to unity at the present time. So the angular-diameter distance is just
\begin{equation}
	\label{eq:ang_dist}
	d_\text{A}=\frac{1}{1+z}\int_0^z \frac{dz'}{H(z')}\,.
\end{equation}
Comparing Eqs. \eqref{eq:lum_dist} and \eqref{eq:ang_dist} one finds
\begin{equation}
	\label{eq:mod_DDR}
	\frac{d_\text{L}}{d_\text{A}}=(1+z)^2 \sqrt{\frac{f_2(0)}{f_2(z)}}\,.
\end{equation}

Deviations from the standard DDR are usually parametrized by the factor $\eta$ as 
\begin{equation}
	\label{eq:par_DDR}
	\frac{d_\text{L}}{d_\text{A}}=(1+z)^2 \eta\,.
\end{equation}
Constraints on the value of $\eta$ are derived from observational data for both $d_\text{A}$ and $d_\text{L}$.
Comparing Eqs. \eqref{eq:mod_DDR} and \eqref{eq:par_DDR} one immediately obtains
\begin{equation}
	\label{eq:eta}
	\eta(z) = \sqrt{\frac{f_2(0)}{f_2(z)}}\,,
\end{equation}
(see also \cite{Minazzoli2014,Hees2014} for a derivation of this result in theories with a NMC between the matter fields and a scalar field). 

If $f_1=R$, like in GR, any choice of the NMC function apart from $f_2 = 1$ would lead to a deviation from the standard $\Lambda$CDM background cosmology and would therefore require a computation of the modified $H(z)$ and $R(z)$ for every different $f_2$ that is probed. However, it is possible to choose a function $f_1$ such that the cosmological background evolution remains the same as in $\Lambda$CDM. In this case, the Hubble factor is simply 
	\begin{equation}
		\label{eq:Hubble}
		H(z)=H_0\left[\Omega_{\text{r},0}(1+z)^4+\Omega_{\text{m},0}(1+z)^3+\Omega_{\Lambda,0}\right]^{1/2} \,,
	\end{equation}
	and the scalar curvature $R$ is given by
	\begin{equation}
		\label{eq:scalar_curv}
		R(z)=3H_0^2\left[\Omega_{\text{m},0}(1+z)^3+4\Omega_{\Lambda,0}\right] \,,
	\end{equation}
	where $H_0$ is the Hubble constant, $\Omega_{\text{r},0}$, $\Omega_{\text{m},0}$ and $\Omega_{\Lambda,0}$ are the radiation, matter and cosmological constant density parameters at present time. The calculation of the appropriate function $f_1$ must be done numerically, by integrating either the MFE \eqref{eq:fried-f1f2-1} or the MRE \eqref{eq:ray-f1f2-1} for $f_1$ with the appropriate initial conditions at $z=0$ (when integrating the MRE, the MFE serves as an additional constraint). Considering that GR is strongly constrained at the present time, the natural choice of initial conditions is $f_1(0)=R(0)$ and $\dot{f}_1(0)=\dot{R}(0)$.

Nevertheless, in the present paper we will consider that significant deviations of $f_2$ from unity are allowed only at relatively low redshift, since CMB and BBN constraints on NMC theories have already constrained $f_2$ to be very close to unity at large redshifts \cite{Avelino2018,Azevedo2018a}. We have verified that the function $f_1(z)$ required for Eqs. \eqref{eq:Hubble} and \eqref{eq:scalar_curv} to be satisfied deviates no more than $3\%$ from the GR prediction $f_1=R$ for $z\lesssim1.5$, for the models investigated in this paper (using the best-fit parameters in Table \ref{tab:results_full_n}).

\section{Methodology and results}\label{sec:results}

In \cite{Martinelli2020}, the authors used Pantheon and BAO data to constrain a parametrization of the DDR deviation of the type 
\begin{equation}
	\label{eq:par_eta_eps}
	\eta(z) = (1+z)^{\epsilon}\,.
\end{equation}
and obtained, for a constant $\epsilon$, $\epsilon=0.013\pm0.029$ at the 68\% credible interval (CI). Here, we use the same datasets and a similar methodology to derive constraints for specific NMC models. We present a brief description of the methodology for completeness, but refer the reader to \cite{Martinelli2020} for a more detailed discussion. 

In general, BAO data provides measurements of the ratio $d_z$ (see, for example, \cite{Beutler2011}), defined as
\begin{equation}
	d_z\equiv \frac{r_\text{s}(z_\text{d})}{D_V(z)} \,,
\end{equation}
where $D_V(z)$ is the volume averaged distance \cite{Eisenstein2005}
\begin{equation}
	D_V(z)=\left[(1+z)^2 d_\text{A}^2(z) \frac{c z}{H(z)}\right]^{1/3} \,,
\end{equation}
and $r_\text{s}(z_\text{d})$ is the comoving sound horizon at the drag epoch. Assuming that the evolution of the Universe is close to $\Lambda$CDM, $r_\text{s}(z_\text{d})$ can be approximated as \cite{Eisenstein1998}
\begin{equation}
	\label{eq:sound_hor}
	r_\text{s}(z_\text{d}) \simeq \frac{44.5 \ln \left(\frac{9.83}{\Omega_{\text{m},0}h^2}\right)} {\sqrt{1+10(\Omega_{\text{b},0}h^2)^{3/4}}} \,,
\end{equation}
where $\Omega_{\text{b},0}$ is the baryon density parameter and $h$ is the dimensionless Hubble constant. Here we shall assume that $\Omega_{\text{b},0}h^2=0.02225$ in agreement with the latest \textit{Planck} release \cite{Aghanim2020}. Notice that the BAO observations are used to estimate $d_\text{A}$, which remains unchanged in NMC theories provided that the evolution of $H(z)$ and $R(z)$ is unchanged with respect to the $\Lambda$CDM model. Thus, BAO data will ultimately provide us with constraints on $H_0$ and $\Omega_{\text{m},0}$. The original datasets that we shall consider in the present paper come from the surveys 6dFGS \cite{Beutler2011}, SDDS \cite{Anderson2014}, BOSS CMASS \cite{Xu2012}, WiggleZ \cite{Blake2012}, MGS \cite{Ross2015}, BOSS DR12 \cite{Gil-Marin2016}, DES \cite{Abbott2019}, Ly-$\alpha$ observations \cite{Blomqvist2019}, SDSS DR14 LRG \cite{Bautista2018} and quasar observations \cite{Ata2018}, but the relevant data is conveniently compiled and combined in Appendix A of \cite{Martinelli2020}.

Likewise, the luminosity distance can be constrained using SnIa data, via measurements of their apparent magnitude
\begin{equation}
	\label{eq:appar_mag}
	m(z)=M_0 +5 \log_{10}\left[\frac{d_\text{L}(z)}{\text{Mpc}}\right]+25 \,,
\end{equation}
or, equivalently,
\begin{equation}
	\label{eq:appar_mag_explicit}
	m(z) = M_0 - 5\log_{10}(H_0)+ 5\log_{10}\left[\eta(z)\hat{d}_\text{L}(z)\right] +25 \,,
\end{equation}
where $M_0$ is the intrinsic magnitude of the supernova and $\hat{d}_\text{L}(z)$ is the GR Hubble constant-free luminosity distance. Note, that the intrinsic magnitude $M_0$ is completely degenerate with the Hubble constant $H_0$, and thus simultaneous constraints on both quantities cannot be derived from SnIa data alone. As per \cite{Martinelli2020}, we use the marginalized likelihood expression from Appendix C in \cite{Conley2011}, which takes into account the marginalization of both $M_0$ and $H_0$, whenever possible.  Likewise, we use the full 1048 point Pantheon compilation from \cite{Scolnic2018}.

For simplicity, we shall consider two NMC models with a single free parameter (the NMC parameter) which is assumed to be a constant in the relevant redshift range ($0<z<1.5$), and assume a flat Universe evolving essentially as $\Lambda$CDM. Since the contribution of radiation to the overall energy density is very small at low redshift we ignore its contribution, and therefore $\Omega_{\Lambda,0}=1-\Omega_{\text{m},0}$.

We use the Markov chain Monte Carlo (MCMC) sampler in the publicly available Python package \texttt{emcee} \cite{Foreman-Mackey2013} to build the posterior likelihoods for the cosmological parameters, $H_0$ and $\Omega_{\text{m},0}$, as well as the NMC parameter, assuming flat priors for all of them. The MCMC chains are then analyzed using the Python package \texttt{GetDist} \cite{Lewis2019}, in order to calculate the marginalized means and CIs, as well as plots of the 2D contours of the resulting distributions.

\subsection{Power Law}
\label{subsec:powerlaw}

Consider a power law NMC function of the type
\begin{equation}
	\label{eq:power_law}
	f_2\propto R^n \,,
\end{equation}
where $n$ is the NMC parameter (GR is recovered when $n=0$). Using Eqs. \eqref{eq:scalar_curv} and \eqref{eq:power_law} in Eq. \eqref{eq:eta}, one obtains
\begin{equation}
	\label{eq:eta_power}
	\eta(z;n,\Omega_{\text{m},0}) = \left[\frac{\Omega_{\text{m},0}+4(1-\Omega_{\text{m},0})}{\Omega_{\text{m},0}(1+z)^3+4(1-\Omega_{\text{m},0})}\right]^{n/2} \,.
\end{equation}

\begin{figure}
	\centering
	\includegraphics[width=\columnwidth]{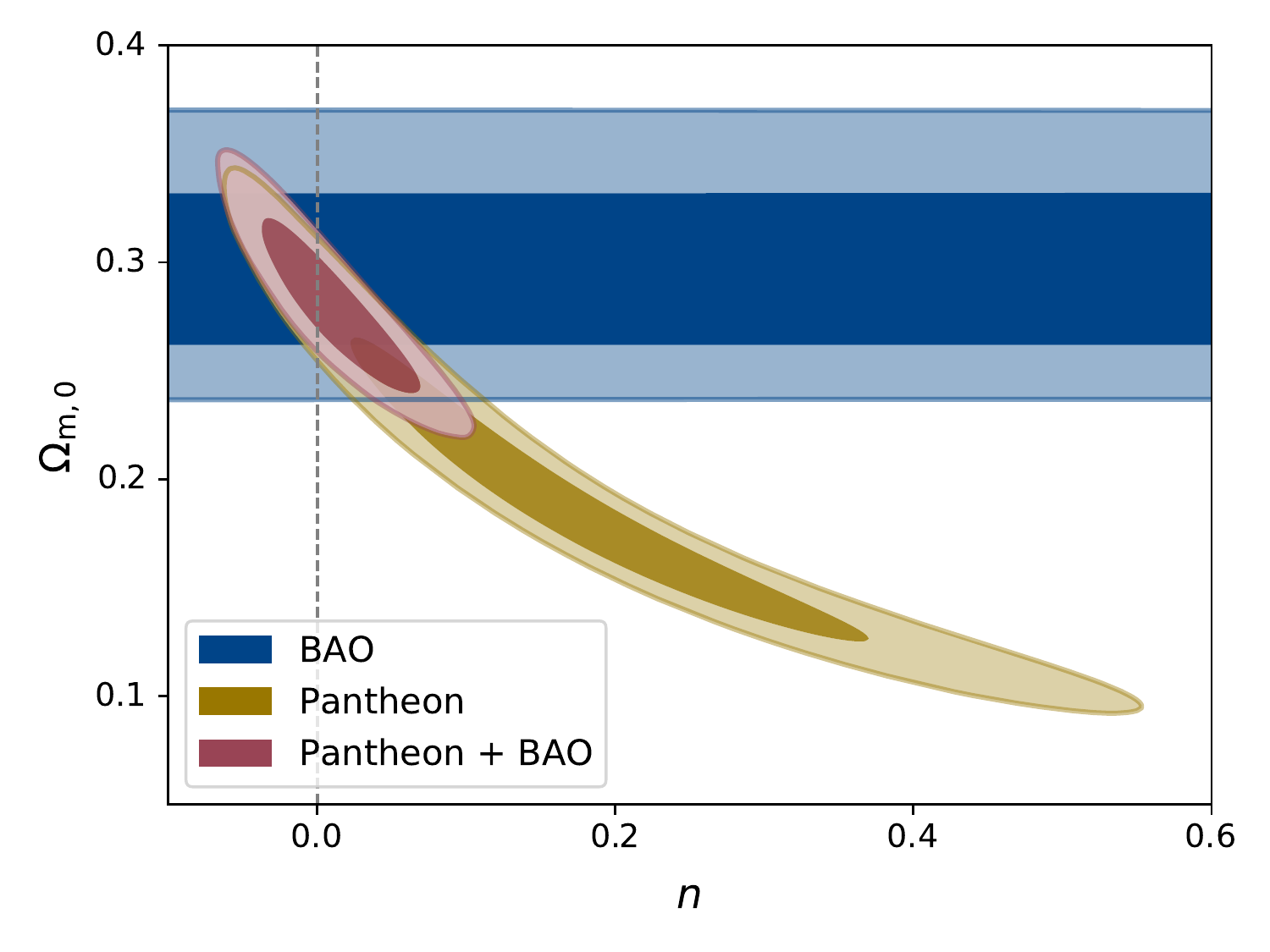}
	\caption{2D contours on the power law parameter $n$ and $\Omega_{\text{m},0}$ using data from BAO (blue), SnIa (yellow) and the combination of the two (red). The darker and lighter concentric regions represent the 68\% and 95\% credible intervals, respectively.}
	\label{fig:n_Om}
\end{figure}

\begin{figure}
	\centering
	\includegraphics[width=\columnwidth]{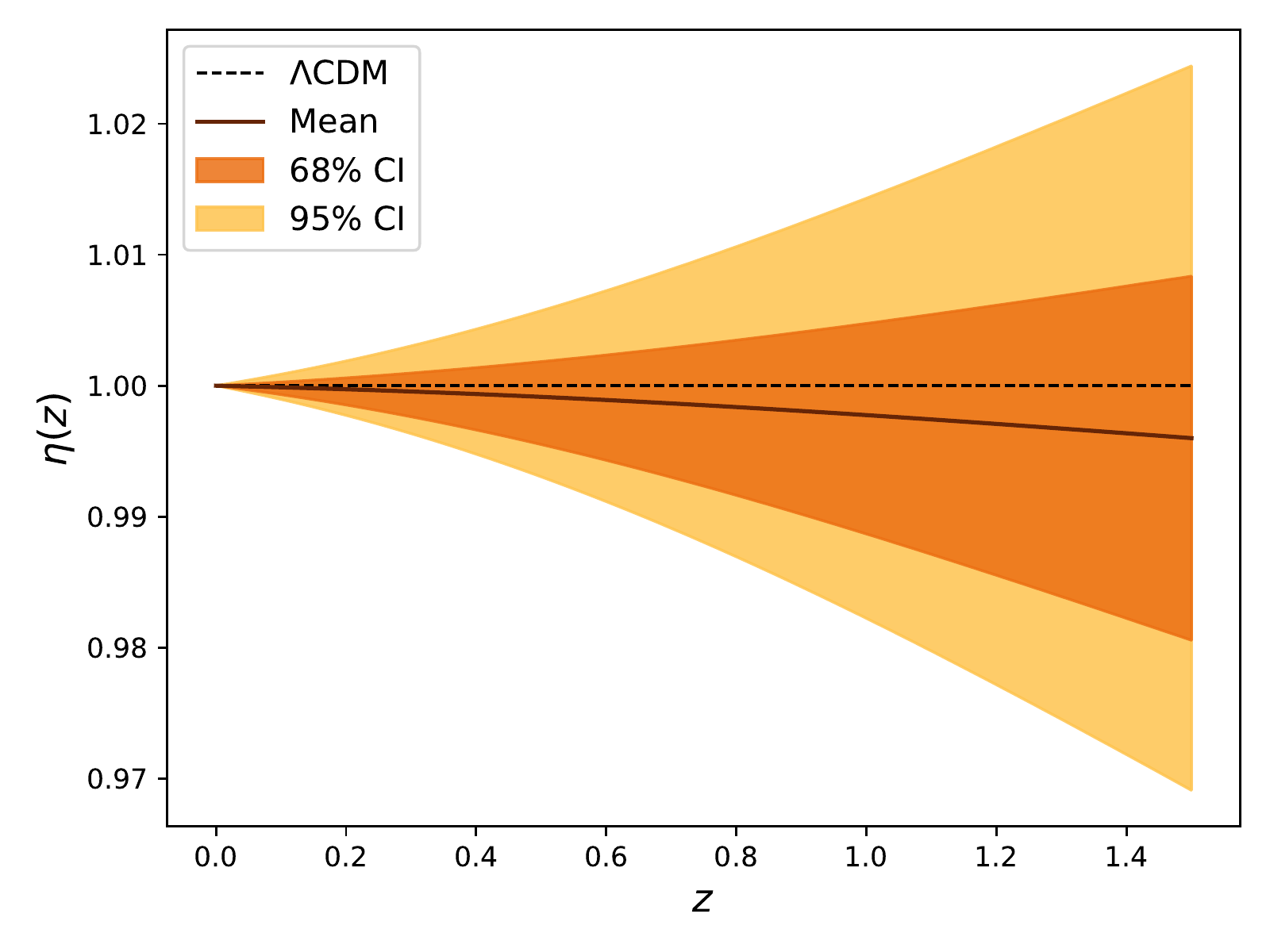}
	\caption{Reconstruction of $\eta(z)$ for the power law NMC model, using the results from the analysis of combined BAO and SnIa data. The dashed line represents the GR prediction $\eta=1$, while the solid red line represents the mean value of $\eta$ at every redshift. The orange (darker) and yellow (lighter) contours represent the 68\% and 95\% credible intervals, respectively.}
	\label{fig:eta_power_law}
\end{figure}

The marginalized 68\% CI results can be found in Table \ref{tab:results}, and the 2D distributions for $n$ and $\Omega_{\text{m},0}$ are displayed in Fig.~\ref{fig:n_Om} (see also the Appendix for the best-fit values, 95\% and 99\% CIs, and the remaining distribution plots). A reconstruction of Eq. \eqref{eq:eta_power} is also shown in Fig.~\ref{fig:eta_power_law}. Note that Eq. (35) implies that in the power law case $\eta(z)$ only depends on the parameters $n$ and $\Omega_{\text{m},0}$, which are not completely degenerate. Therefore,  SnIa data alone is able to constrain both of these parameters. However, since BAO data constrains both $H_0$ and $\Omega_{\text{m},0}$, we are able to combine the two datasets to significantly improve the constraints on $n$ and $\Omega_{\text{m},0}$.

The combined SnIa and BAO datasets constrain the NMC parameter to $n=0.013\pm 0.035$ (68\% CI). While this constraint falls short of the ones previously obtained from the black-body spectrum of the CMB, $|n|\lesssim \text{few}\times 10^{-6}$, or from BBN, $-0.002<n<0.003$ \cite{Azevedo2018a}, the present results are complementary as they more directly constrain the value of $n$ at much smaller redshifts. Note that this rules out NMC models designed to mimic dark matter, as these would require a power law with exponent in the range $-1\leq n \leq -1/7$ in order to explain the observed galactic rotation curves \cite{Bertolami2012,Silva2018}, in accordance with previous results \cite{Azevedo2018a}.

\begin{table}
	\caption{Mean values and marginalized 68\% CI limits obtained from currently available data on the cosmological parameters $\Omega_{\text{m},0}$ and $H_0$ (in units of km s$^{-1}$ Mpc$^{−1}$) and on the NMC parameters $n$ (dimensionless) and $\beta$ (in units of km$^{-2}$ s$^2$ Mpc$^2$).}
	\label{tab:results}
	\begin{tabular} { lc|c|c}		
		\hline\hline
		Parameter & Probe  & power law & Exponential \\
		\hline
		& BAO &  $66.8^{+1.2}_{-1.4}$      &  $66.8^{+1.2}_{-1.4}$       \\
		{\boldmath$H_0$} & SnIa &  unconstrained      &  unconstrained       \\
		& SnIa+BAO &  $66.1\pm 1.2$      &  $65.7\pm 1.0$   \\
		\hline                 
		& BAO & $0.300^{+0.027}_{-0.035}$& $0.300^{+0.027}_{-0.035}$ \\		
		{\boldmath$\Omega_{\text{m},0}$} & SnIa & $0.191^{+0.037}_{-0.061}$  & unconstrained \\
		& SnIa+BAO & $0.279^{+0.024}_{-0.030}$  & $0.268\pm 0.019$   \\
		\hline		
		& BAO & unconstrained   &  \----       \\
		{\boldmath$n$} & SnIa & $0.184^{+0.092}_{-0.15}$   & \----  \\
		& SnIa+BAO & $0.013\pm 0.035$   &   \----      \\
		\hline
		& BAO & \----  &  unconstrained      \\
		{\boldmath$\beta $} & SnIa & \----  &  unconstrained       \\
		& SnIa+BAO &\----& $\left(1.24^{+0.97}_{-1.2}\right)\cdot10^{-6}$ \\
		\hline \hline
	\end{tabular}	
\end{table}

\subsection{Exponential}
\label{sec:exp}

\begin{figure}[!]
	\centering
	\includegraphics[width=\columnwidth]{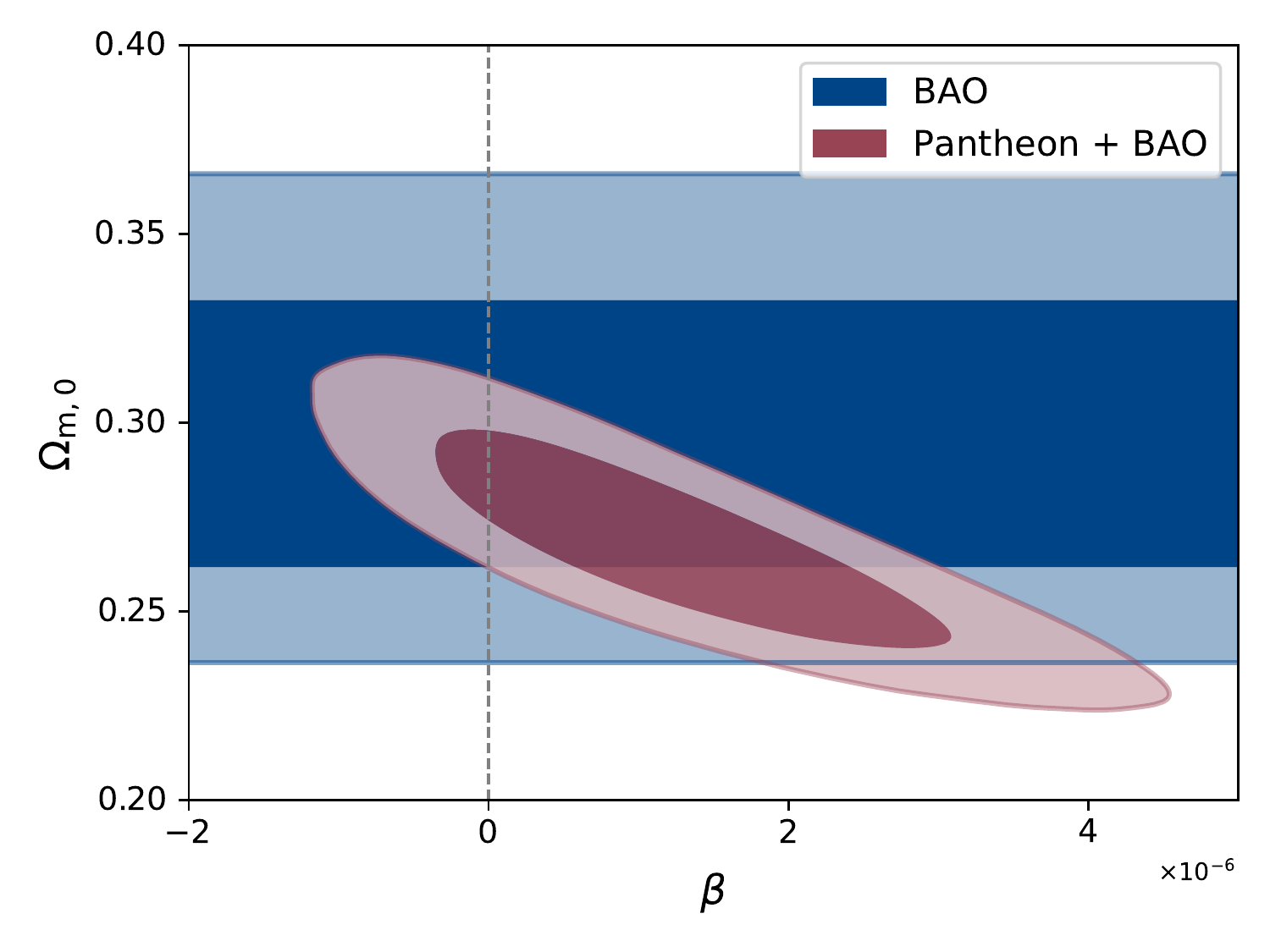}
	\caption{2D contours on the exponential parameter $\beta$ and $\Omega_{\text{m},0}$ using data from BAO (blue) and the combination of the SnIa and BAO (red). The darker and lighter concentric regions represent the 68\% and 95\% credible intervals, respectively.}
	\label{fig:beta_Om}
\end{figure}

\begin{figure}[!]
	\centering
	\includegraphics[width=\columnwidth]{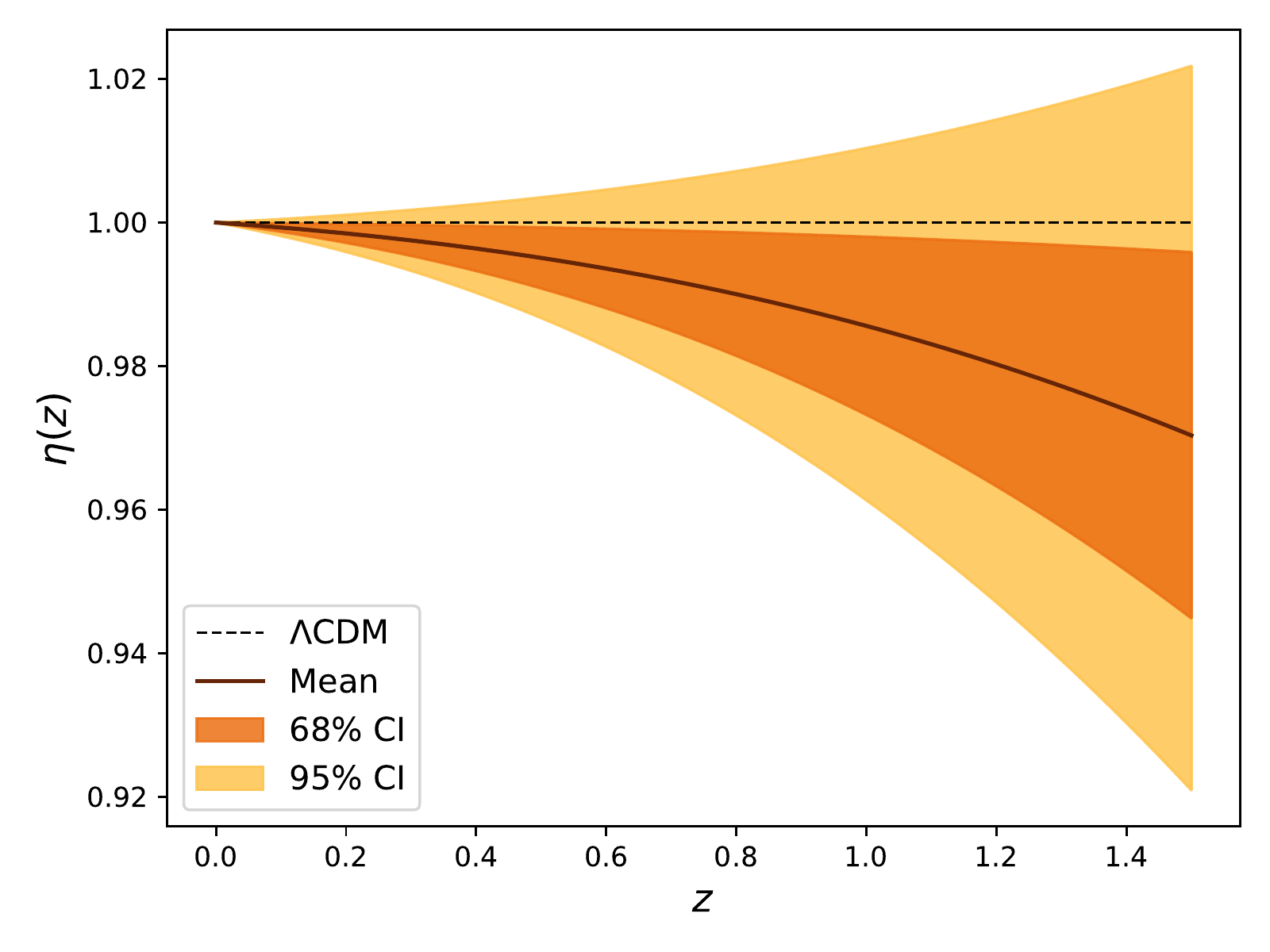}
	\caption{Reconstruction of $\eta(z)$ for the exponential NMC model, using the results from the analysis of combined BAO and SnIa data. The dashed line represents the GR prediction $\eta=1$, while the solid red line represents the mean value of $\eta$ at every redshift. The orange (darker) and yellow (lighter) contours represent the 68\% and 95\% credible intervals, respectively.}
	\label{fig:eta_exp}
\end{figure}

Consider now an exponential NMC function,
\begin{equation}
	\label{eq:exp_f2}
	f_2\propto e^{\beta R} \,,
\end{equation}
where $\beta$ is the NMC parameter, with dimensions of $R^{-1}$ (GR is recovered when $\beta=0$). Using Eqs. \eqref{eq:scalar_curv} and \eqref{eq:exp_f2} in Eq. \eqref{eq:eta}, one obtains
\begin{equation}
	\label{eq:eta_exp}
	\eta(z;\beta,\Omega_{\text{m},0}, H_0) = \exp\left[\frac{3}{2}\beta H_0^2 \Omega_{\text{m},0} (1-(1+z)^3)\right] \,.
\end{equation}
Note that $\eta$ now depends on all three free parameters, $\beta$, $\Omega_{\text{m},0}$ and $H_0$. Furthermore, since $H_0$ is now also degenerate with $\beta$ and $\Omega_{\text{m},0}$, we can no longer analytically marginalize over $H_0$, and SnIa data alone cannot be used to derive useful constraints on any of these parameters. By combining the BAO and SnIa datasets, however, one is able to break this degeneracy, and derive constraints on all three parameters. The marginalized results can be found in Table \ref{tab:results}, and the 2D distributions for $\beta$ and $\Omega_{\text{m},0}$ can be found in Fig.~\ref{fig:beta_Om} (see also the Appendix for the best-fit values, $95\%$ and $99\%$ CIs, and the remaining distribution plots). A reconstruction of Eq. \eqref{eq:eta_exp} is also shown in Fig.~\ref{fig:eta_exp}.

The combined SnIa and BAO datasets constrain the NMC parameter to $\beta=\left(1.24^{+0.97}_{-1.2}\right)\cdot10^{-6}$ (68\% CI), in units of km$^{-2}$ s$^2$ Mpc$^2$. Once again this result complements the one found for the same function using the method presented in \cite{Azevedo2018a} for the variation of the baryon to photon ratio, $|\beta|\lesssim10^{-28}$, as they constrain the same parameter in significantly different redshift ranges. Also notice that while the marginalized results do not contain the GR limit $\beta=0$ at the 68\% CI, that limit is contained in both the marginalized 95\% CI, $\beta=\left(1.2^{+2.2}_{-2.1}\right)\cdot10^{-6}$, and the 2D 68\% credible region.

Future observations from LSST and the Euclid DESIRE survey \cite{Laureijs2011,Astier2014} will provide more data points in the range $z\in[0.1, 1.6]$. For the $\epsilon$ parametrization used in \cite{Martinelli2020}, this will result in an improvement on the constraint of about one order of magnitude. If this is the case, one could expect a corresponding improvement on the NMC parameter constraints. Third-generation gravitational wave observatories will also be able to provide data points at even higher redshift (up to $z\sim5$), which will serve as  independent and complementary data for the same purpose \cite{Cai2017}.

\section{Conclusions}
\label{sec:conc}

In this work, we have shown that the relation between the luminosity and angular-diameter distances is modified in the presence of a NMC between gravity and the matter fields, due to the change in energy and momentum that photons experience as space expands. We have used the most recent SnIa and BAO data to further constrain these $f(R)$-inspired NMC theories in a redshift range closer to the present time (up to $z\sim1.5$), which had previously not been explored in the context of NMC gravity. We find that the deviations from GR are expected to be small, in accordance with previous results. These new constraints also reinforce that NMC gravity cannot be used to explain the observed galactic rotation curves in lieu of dark matter.

\begin{acknowledgments}	
	The authors would like to thank Dr. Paul Tol for developing the colorblind-friendly color schemes used in this work \cite{Tol}. R.P.L.A. was supported by the Funda{\c c}\~ao para a Ci\^encia e Tecnologia (FCT, Portugal) Grant No. SFRH/BD/132546/2017. Funding of this work has also been provided by FCT through national funds (PTDC/FIS-PAR/31938/2017) and by FEDER—Fundo Europeu de Desenvolvimento Regional through COMPETE2020 - Programa Operacional Competitividade e Internacionaliza{\c c}\~ao (POCI-01-0145-FEDER-031938), and through the research Grants No. UIDB/04434/2020 and No. UIDP/04434/2020.
\end{acknowledgments}
\appendix*
\section{Full marginalized result table and distribution plots}

\begin{figure}[H]
	\centering
	\includegraphics[width=\columnwidth]{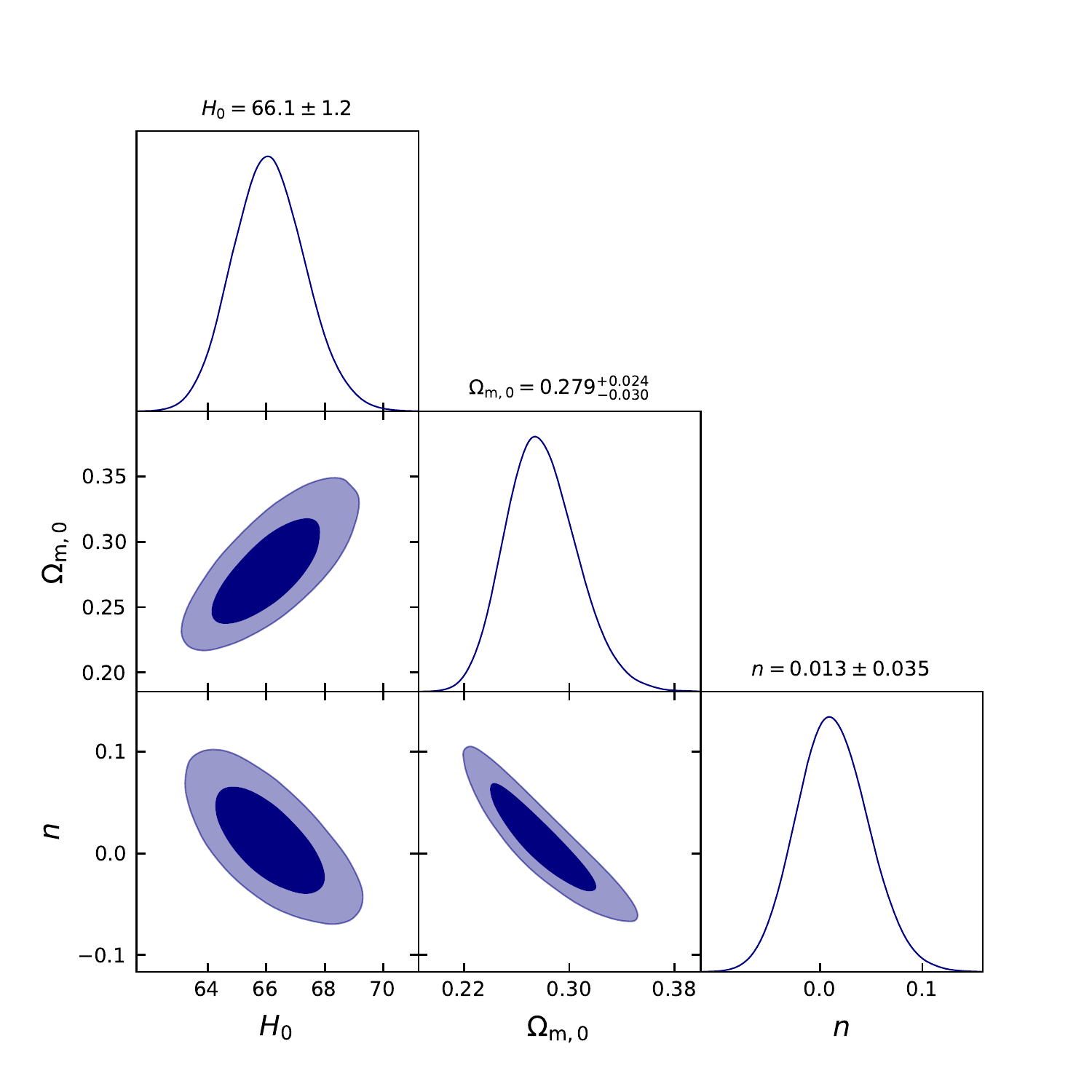}
	\caption{Constraints on the power law parameter $n$, $H_0$ and $\Omega_{\text{m},0}$ using combined data from BAO and SnIa. The darker and lighter regions represent the 68\% and 95\% credible intervals, respectively.}
	\label{fig:tri_n_Tot}
\end{figure}

\begin{figure}[H]
	\centering
	\includegraphics[width=\columnwidth]{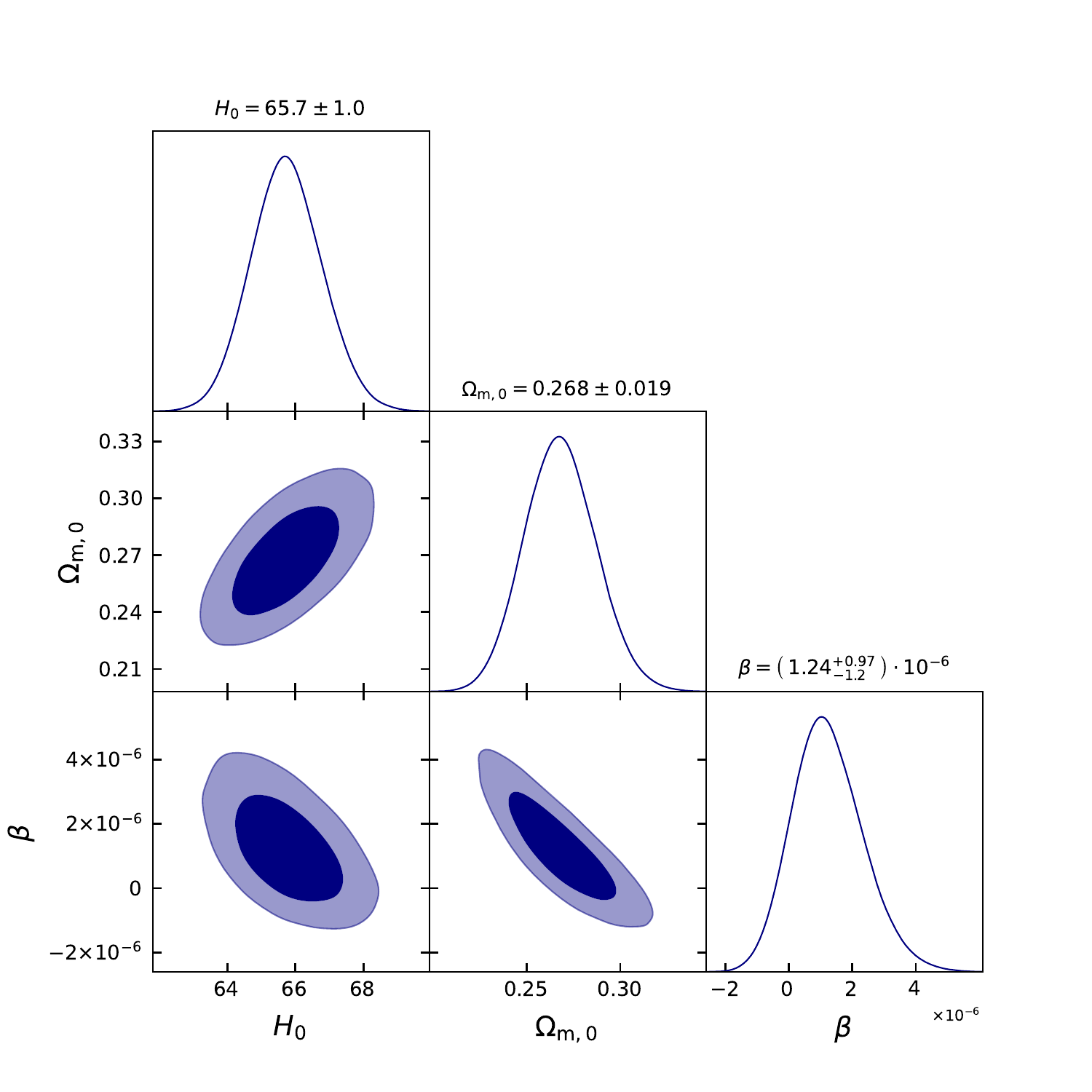}
	\caption{Constraints on the exponential parameter $\beta$, $H_0$ and $\Omega_{\text{m},0}$ using combined data from BAO and SnIa. The darker and lighter regions represent the 68\% and 95\% credible intervals, respectively.}
	\label{fig:tri_beta_Tot}
\end{figure}

The distributions of the cosmological parameters, $H_0$ and $\Omega_{\text{m},0}$, and the NMC parameters, $n$ and $\beta$, for the combined SnIa and BAO datasets can be found in Figs. \ref{fig:tri_n_Tot} and \ref{fig:tri_beta_Tot} for the power law and exponential functions, respectively. In Table \ref{tab:results_full_n} we have also included the best-fit values for each parameter/data-set pair, as well as the marginalized means, 68\%, 95\% and 99\% CIs, since the distributions are not perfectly Gaussian.

\begin{table*}
	\caption{Best-fit values and marginalized means, 68\%, 95\% and 99\% CI limits obtained from currently available data on the cosmological parameters $\Omega_{\text{m},0}$ and $H_0$ (in units of km s$^{-1}$ Mpc$^{−1}$) and on the NMC parameters $n$ (dimensionless) and $\beta$ (in units of km$^{-2}$ s$^2$ Mpc$^2$).}
	\label{tab:results_full_n}
	\begin{tabular}{lc|ccccc|ccccc}
		\hline\hline
		&          & \multicolumn{5}{c|}{power law}                                                          & \multicolumn{5}{c}{Exponential}                                                         \\
		Parameter                        & Probe    & Best fit & Mean    & 68\%                 & 95\%                 & 99\%                 & Best fit & Mean    & 68\%                 & 95\%                 & 99\%                 \\ \hline
		& BAO      & $66.4$   & $66.8$  & $^{+1.2}_{-1.4}$     & $^{+2.7}_{-2.5}$     & $^{+3.7}_{-3.2}$     & $66.4$   & $66.8$  & $^{+1.2}_{-1.4}$     & $^{+2.7}_{-2.5}$     & $^{+3.7}_{-3.2}$     \\
		{\boldmath$H_0$}                 & SnIa     & \multicolumn{5}{c|}{unconstrained}                                                      & \multicolumn{5}{c}{unconstrained}                                                       \\
		& SnIa+BAO & $66.0$   & $66.1$  & $\pm 1.2$            & $^{+2.5}_{-2.3}$     & $^{+3.4}_{-3.0}$     & $65.7$   & $65.7$  & $\pm 1.0$            & $^{+2.1}_{-2.0}$     & $^{+3.4}_{-3.0}$     \\ \hline
		& BAO      & $0.291$  & $0.300$ & $^{+0.027}_{-0.035}$ & $^{+0.064}_{-0.060}$ & $^{+0.094}_{-0.071}$ & $0.291$  & $0.300$ & $^{+0.027}_{-0.035}$ & $^{+0.064}_{-0.060}$ & $^{+0.094}_{-0.071}$ \\
		{\boldmath$\Omega_{\text{m},0}$} & SnIa     & $0.181$  & $0.191$ & $^{+0.037}_{-0.061}$ & $^{+0.11}_{-0.094}$  & $^{+0.18}_{-0.10}$   & \multicolumn{5}{c}{unconstrained}                                                       \\
		& SnIa+BAO & $0.276$  & $0.279$ & $^{+0.024}_{-0.030}$ & $^{+0.054}_{-0.052}$ & $^{+0.079}_{-0.062}$ & $0.268$  & $0.268$ & $\pm0.019$           & $^{+0.038}_{-0.036}$ & $^{+0.052}_{-0.046}$ \\ \hline
		& BAO      & \multicolumn{5}{c|}{unconstrained}                                                      & \multicolumn{5}{c}{\----}                                                               \\
		{\boldmath$n$}                   & SnIa     & $0.178$  & $0.184$ & $^{+0.092}_{-0.15}$  & $^{+0.26}_{-0.24}$   & $^{+0.45}_{-0.26}$   & \multicolumn{5}{c}{\----}                                                               \\
		& SnIa+BAO & $0.014$  & $0.013$ & $\pm 0.035$          & $^{+0.071}_{-0.066}$ & $^{+0.097}_{-0.085}$ & \multicolumn{5}{c}{\----}                                                               \\ \hline
		& BAO      & \multicolumn{5}{c|}{\----}                                                              & \multicolumn{5}{c}{unconstrained}                                                       \\
		{\boldmath$\beta\cdot10^6$}      & SnIa     & \multicolumn{5}{c|}{\----}                                                              & \multicolumn{5}{c}{unconstrained}                                                       \\
		& SnIa+BAO & \multicolumn{5}{c|}{\----}                                                              & $1.18$   & $1.24$  & $^{+0.97}_{-1.2}$    & $^{+2.2}_{-2.1}$     & $^{+3.3}_{-2.5}$     \\ \hline\hline
	\end{tabular}
\end{table*}

\bibliography{NMC_distance.bib}

\end{document}